\documentclass[aps,prd,twocolumn,superscriptaddress,nofootinbib]{revtex4-1}

\usepackage[dvipsnames]{xcolor}
\usepackage{empheq}
\usepackage{gensymb}
\usepackage{graphicx}
\usepackage{dcolumn}
\usepackage{bm}
\usepackage{hyperref}
\usepackage{lineno}

\usepackage[abs]{overpic}

\hypersetup{
    pdfnewwindow=true,      
    colorlinks=true,       
    linkcolor=blue,          
    citecolor=blue,        
    filecolor=blue,      
    urlcolor=blue           
}

\newcommand{\Minn}{MINER$\nu$A}
\newcommand{\Min}{\Minn~}

\newcommand{\genien}{GENIE}
\newcommand{\genie}{\genien~}

\newcommand{\gibuun}{GiBUU}
\newcommand{\gibuu}{\gibuun~}
\newcommand{\bvern}{2019}
\newcommand{\bver}{\bvern~}

\newcommand{\mnvgvern}{2.8.4}
\newcommand{\mnvgver}{\mnvgvern~}

\newcommand{\nuwron}{NuWro}
\newcommand{\nuwro}{\nuwron~}

\newcommand{\nvern}{19.02}
\newcommand{\nver}{\nvern~}

\newcommand{\pythian}{\textsc{pythia6}}
\newcommand{\pythia}{\pythian~}

\newcommand{\geantn}{\textsc{Geant4}}
\newcommand{\geant}{\geantn~}

\newcommand{\gfvern}{\textsc{9.4.2}}

\newcommand{\minosn}{MINOS}
\newcommand{\minos}{\minosn~}

\newcommand{\dunen}{DUNE}

\newcommand{\pizn}{$\pi^0$}
\newcommand{\piz}{\pizn~}
\newcommand{\hadron}{\textrm{h}}
\newcommand{\proton}{\textrm{p}}
\newcommand{\neutron}{\textrm{n}}

\newcommand{\numun}{$\nu_\mu$}
\newcommand{\numu}{\numun~}

\newcommand{\nn}{neutrino}
\newcommand{\n}{\nn~}

\newcommand{\Nn}{Neutrino}
\newcommand{\N}{\Nn~}

\newcommand{\nsn}{\nn s}
\newcommand{\ns}{\nsn~}

\newcommand{\cpn}{$\mathcal{CP}$}

\newcommand{\dpttn}{$\delta p_\textrm{TT}$}
\newcommand{\dptt}{\dpttn~}

\newcommand{\dptn}{$\delta p_\textrm{T}$}
\newcommand{\dpt}{\dptn~}

\newcommand{\dalphatn}{$\delta \alpha_\textrm{T}$}
\newcommand{\dalphat}{\dalphatn~}

\newcommand{\pnn}{$p_\textrm{n}$}
\newcommand{\pN}{p_\textrm{n}}
\newcommand{\pn}{\pnn~}

\newcommand{\fsin}{FSI}
\newcommand{\fsi}{\fsin~}

\newcommand{\mevn}{MeV}
\newcommand{\mev}{\mevn~}

\newcommand{\gevn}{GeV}

\newcommand{\ntargets}{$3.12\times 10^{30}$}

\newcommand{\topon}{$1\mu^-N\textrm{p}M\pi^0~(N,M>0)$}
\newcommand{\topo}{\topon~}

\newcommand{\sigwidth}{\columnwidth}

\begin{document}

\title{Probing Nuclear Effects with Neutrino-Induced Charged-Current  Neutral Pion Production}  

\newcommand{\Rutgers}{Rutgers, The State University of New Jersey, Piscataway, New Jersey 08854, USA}
\newcommand{\Hampton}{Hampton University, Dept. of Physics, Hampton, VA 23668, USA}
\newcommand{\Dortmund}{Institute of Physics, Dortmund University, 44221, Germany }
\newcommand{\Otterbein}{Department of Physics, Otterbein University, 1 South Grove Street, Westerville, OH, 43081 USA}
\newcommand{\JMU}{James Madison University, Harrisonburg, Virginia 22807, USA}
\newcommand{\Florida}{University of Florida, Department of Physics, Gainesville, FL 32611}
\newcommand{\UCIrvine}{Department of Physics and Astronomy, University of California, Irvine, Irvine, California 92697-4575, USA}
\newcommand{\CBPF}{Centro Brasileiro de Pesquisas F\'{i}sicas, Rua Dr. Xavier Sigaud 150, Urca, Rio de Janeiro, Rio de Janeiro, 22290-180, Brazil}
\newcommand{\PUCP}{Secci\'{o}n F\'{i}sica, Departamento de Ciencias, Pontificia Universidad Cat\'{o}lica del Per\'{u}, Apartado 1761, Lima, Per\'{u}}
\newcommand{\INRM}{Institute for Nuclear Research of the Russian Academy of Sciences, 117312 Moscow, Russia}
\newcommand{\Jlab}{Jefferson Lab, 12000 Jefferson Avenue, Newport News, VA 23606, USA}
\newcommand{\Pittsburgh}{Department of Physics and Astronomy, University of Pittsburgh, Pittsburgh, Pennsylvania 15260, USA}
\newcommand{\Guanajuato}{Campus Le\'{o}n y Campus Guanajuato, Universidad de Guanajuato, Lascurain de Retana No. 5, Colonia Centro, Guanajuato 36000, Guanajuato M\'{e}xico.}
\newcommand{\Athens}{Department of Physics, University of Athens, GR-15771 Athens, Greece}
\newcommand{\Tufts}{Physics Department, Tufts University, Medford, Massachusetts 02155, USA}
\newcommand{\WM}{Department of Physics, College of William \& Mary, Williamsburg, Virginia 23187, USA}
\newcommand{\FNAL}{Fermi National Accelerator Laboratory, Batavia, Illinois 60510, USA}
\newcommand{\Purdue}{Department of Chemistry and Physics, Purdue University Calumet, Hammond, Indiana 46323, USA}
\newcommand{\MCLA}{Massachusetts College of Liberal Arts, 375 Church Street, North Adams, MA 01247}
\newcommand{\UMD}{Department of Physics, University of Minnesota -- Duluth, Duluth, Minnesota 55812, USA}
\newcommand{\Northwestern}{Northwestern University, Evanston, Illinois 60208}
\newcommand{\UNI}{Universidad Nacional de Ingenier\'{i}a, Apartado 31139, Lima, Per\'{u}}
\newcommand{\Rochester}{University of Rochester, Rochester, New York 14627 USA}
\newcommand{\Austin}{Department of Physics, University of Texas, 1 University Station, Austin, Texas 78712, USA}
\newcommand{\USM}{Departamento de F\'{i}sica, Universidad T\'{e}cnica Federico Santa Mar\'{i}a, Avenida Espa\~{n}a 1680 Casilla 110-V, Valpara\'{i}so, Chile}
\newcommand{\Geneva}{University of Geneva, 1211 Geneva 4, Switzerland}
\newcommand{\Chicago}{Enrico Fermi Institute, University of Chicago, Chicago, IL 60637 USA}
\newcommand{\hired}{}
\newcommand{\OregonState}{Department of Physics, Oregon State University, Corvallis, Oregon 97331, USA}
\newcommand{\oxford}{Oxford University, Department of Physics, Oxford, OX1 3PJ United Kingdom}
\newcommand{\umiss}{University of Mississippi, Oxford, Mississippi 38677, USA}
\newcommand{\upenn}{Department of Physics and Astronomy, University of Pennsylvania, Philadelphia, PA 19104}
\newcommand{\AMU}{AMU Campus, Aligarh, Uttar Pradesh 202001, India}
\newcommand{\wroclaw}{University of Wroclaw, plac Uniwersytecki 1, 50-137 Wrocław, Poland}
\newcommand{\Mohali}{Department of Physical Sciences, IISER Mohali, Knowledge City, SAS Nagar, Mohali - 140306, Punjab, India}
\newcommand{\CINVESTAV}{Departamento de Fisica Col. San Pedro Zacatenco, 07360 Mexico, DF, Av. Instituto Politécnico Nacional, Mexico}
\newcommand{\york}{York University, Department of Physics and Astronomy, Toronto, Ontario, M3J 1P3 Canada}
\newcommand{\mateusfcarneiroThanks}{Now at Brookhaven National Laboratory}

\author{D.~Coplowe}                       \affiliation{\oxford}
\author{O.~Altinok}                       \affiliation{\Tufts}
\author{Z.~~Ahmad~Dar}                    \affiliation{\AMU}
\author{F.~Akbar}                         \affiliation{\AMU}

\author{D.A.~Andrade}                     \affiliation{\Guanajuato}

\author{G.D.~Barr} \affiliation{\oxford}

\author{A.~Bashyal}                       \affiliation{\OregonState}
\author{A.~Bercellie}                     \affiliation{\Rochester}
\author{M.~Betancourt}                    \affiliation{\FNAL}
\author{A.~Bodek}                         \affiliation{\Rochester}
\author{A.~Bravar}                        \affiliation{\Geneva}
\author{H.~Budd}                          \affiliation{\Rochester}
\author{G.~Caceres}                       \affiliation{\CBPF}
\author{T.~Cai}                           \affiliation{\Rochester}
\author{M.F.~Carneiro}\thanks{\mateusfcarneiroThanks}  \affiliation{\OregonState}  \affiliation{\CBPF}

\author{H.~da~Motta}                      \affiliation{\CBPF}
\author{S.A.~Dytman}                      \affiliation{\Pittsburgh}
\author{G.A.~D\'{i}az~}                   \affiliation{\Rochester}  \affiliation{\PUCP}
\author{J.~Felix}                         \affiliation{\Guanajuato}
\author{L.~Fields}                        \affiliation{\FNAL}
\author{A.~Filkins}                       \affiliation{\WM}
\author{R.~Fine}                          \affiliation{\Rochester}
\author{A.M.~Gago}                        \affiliation{\PUCP}
\author{H.~Gallagher}                     \affiliation{\Tufts}
\author{A.~Ghosh}                         \affiliation{\USM}  \affiliation{\CBPF}
\author{R.~Gran}                          \affiliation{\UMD}
\author{D.A.~Harris}                      \affiliation{\york}  \affiliation{\FNAL}
\author{S.~Henry}                         \affiliation{\Rochester}
\author{S.~Jena}                          \affiliation{\Mohali}
\author{J.~Kleykamp}                      \affiliation{\Rochester}
\author{M.~Kordosky}                      \affiliation{\WM}
\author{D.~Last}                          \affiliation{\upenn}
\author{T.~Le}                            \affiliation{\Tufts}  \affiliation{\Rutgers}
\author{A.~Lozano}                        \affiliation{\CBPF}
\author{X.-G.~Lu} \email[Corresponding author:] {xianguo.lu@physics.ox.ac.uk}    \affiliation{\oxford}
\author{E.~Maher}                         \affiliation{\MCLA}
\author{S.~Manly}                         \affiliation{\Rochester}
\author{W.A.~Mann}                        \affiliation{\Tufts}
\author{C.~Mauger}                        \affiliation{\upenn}
\author{K.S.~McFarland}                   \affiliation{\Rochester}
\author{B.~Messerly}                      \affiliation{\Pittsburgh}
\author{J.~Miller}                        \affiliation{\USM}
\author{J.G.~Morf\'{i}n}                  \affiliation{\FNAL}
\author{D.~Naples}                        \affiliation{\Pittsburgh}
\author{J.K.~Nelson}                      \affiliation{\WM}
\author{C.~Nguyen}                        \affiliation{\Florida}
\author{A.~Norrick}                       \affiliation{\WM}
\author{A.~Olivier}                       \affiliation{\Rochester}
\author{V.~Paolone}                       \affiliation{\Pittsburgh}
\author{G.N.~Perdue}                      \affiliation{\FNAL}  \affiliation{\Rochester}
\author{M.A.~Ram\'{i}rez}                 \affiliation{\Guanajuato}
\author{R.D.~Ransome}                     \affiliation{\Rutgers}
\author{H.~Ray}                           \affiliation{\Florida}
\author{P.A.~Rodrigues}                   \affiliation{\oxford} \affiliation{\umiss}  \affiliation{\Rochester}
\author{D.~Ruterbories}                   \affiliation{\Rochester}
\author{H.~Schellman}                     \affiliation{\OregonState}
\author{J.T.~Sobczyk}                     \affiliation{\wroclaw}
\author{C.J.~Solano~Salinas}              \affiliation{\UNI}
\author{H.~Su}                            \affiliation{\Pittsburgh}
\author{M.~Sultana}                       \affiliation{\Rochester}
\author{V.S.~Syrotenko}                   \affiliation{\Tufts}
\author{E.~Valencia}                      \affiliation{\WM}  \affiliation{\Guanajuato}
\author{D.~Wark}                          \affiliation{\oxford}
\author{A.~Weber}                         \affiliation{\oxford}
\author{M.Wospakrik}                      \affiliation{\Florida}
\author{C.~Wret}                          \affiliation{\Rochester}
\author{B.~Yaeggy}                        \affiliation{\USM}
\author{L.~Zazueta}                       \affiliation{\WM}

\collaboration{The MINER$\nu$A Collaboration}\ \noaffiliation
\date{\today}

\begin{abstract}

We study neutrino-induced charged-current (CC) $\pi^0$ production on carbon nuclei using events with fully imaged final-state proton-$\pi^0$ systems. Novel use of final-state correlations based on transverse kinematic imbalance enable the first measurements of the struck nucleon's Fermi motion, of the intranuclear momentum transfer (IMT) dynamics, and of the final-state hadronic momentum configuration in neutrino pion production.  
Event distributions are presented for i)  the momenta of neutrino-struck neutrons below the Fermi surface, ii) the direction of missing transverse momentum characterizing the strength of IMT, and iii) proton-pion momentum imbalance with respect to the lepton scattering plane. 
The observed Fermi motion and IMT strength are compared to the previous \Min measurement of neutrino CC quasielastic-like production. The measured shapes and absolute rates of these distributions, as well as the cross-section asymmetries show tensions with predictions from current neutrino generator models.

\end{abstract}

\maketitle

\section{Introduction}

In high-statistics \n oscillation experiments,  the measurement precision of the fundamental properties of \ns is becoming limited by  knowledge of \nn-nucleus interactions~\cite{Alvarez-Ruso:2017oui}.
The nuclear medium introduces deviations from free-nucleon scattering that are poorly known and are leading sources of systematic uncertainty for measurements of the \cpn-violating phase~\cite{Acero:2019ksn, Abe:2019vii}.
Among the prominent interaction modes produced when studying \n flavor transformation are final states containing pions. For future experiments such as \dunen~\cite{Abi:2020evt} and HyperKamiokande~\cite{Abe:2018uyc} to achieve their designed sensitivity it is imperative to assess the various nuclear effects involved in pion production.
  
\begin{figure}[!htb]
\centering
\includegraphics[width=0.8\columnwidth]{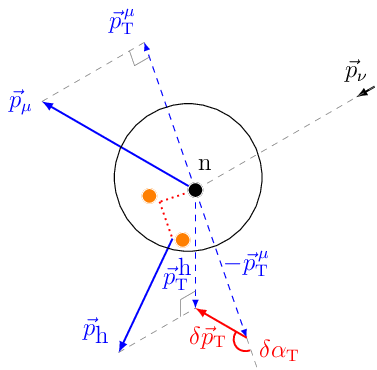}
\caption{Schematic diagram of the single-transverse kinematics~\cite{Lu:2015tcr}. }\label{fig:stkidiagram}
\end{figure}

In previous studies, \Min measured the final-state correlations in neutrino charged-current (CC) quasielastic (QE)-like interactions on carbon~\cite{Lu:2018stk, Cai:2019hpx}. These correlations are based on transverse kinematic imbalance (TKI) that helps precisely identify intranuclear dynamics~\cite{Lu:2015tcr, Furmanski:2016wqo, Abe:2018pwo, Dolan:2018sbb, Lu:2018stk, Dolan:2018zye, Lu:2019nmf, Harewood:2019rzy, Cai:2019jzk, Cai:2019hpx} or the absence thereof~\cite{Lu:2015hea, Duyang:2018lpe, Duyang:2019prb, Munteanu:2019llq, Hamacher-Baumann:2020ogq}. 
The previously measured imbalance is between the CC muon and the final-state proton; the present work studies the following reaction including additional neutral pions in the final state:
\begin{equation}
	\nu_\mu + \textrm{C} \to \mu^- +\underbrace{~\proton + \pi^0}_{\hadron}~+\textrm{X},
	\label{eq:topo}
\end{equation}
where X is a final-state hadronic system consisting of the nuclear remnant with possible additional protons and neutral pions but without other mesons. 
The only difference in the kinematics here is to replace the previous proton final state by $\hadron$ [Eq.~(\ref{eq:topo})] that is the combination of the proton and neutral pion. Therefore, the transverse boosting angle, \dalphatn~\cite{Lu:2015tcr}, and the emulated nucleon momentum, \pnn~\cite{Furmanski:2016wqo, Lu:2019nmf}, are defined similarly as in Ref.~\cite{Lu:2018stk}:
\begin{empheq}[left=\empheqlbrace]{align}
\vec{p}_\hadron &= \vec{p}_\proton + \vec{p}_{\pi^0},\\
\delta\vec{p}_\textrm{T} 	&= \vec{p}_\textrm{T}^{\,\mu} + \vec{p}_\textrm{T}^{\,\hadron},\\
\delta\alpha_\textrm{T}&=\arccos\frac{-\vec{p}_\textrm{T}^{\,\mu}\cdot\delta\vec{p}_\textrm{T}}{p_\textrm{T}^{\,\mu}\delta p_\textrm{T}},\label{eq:dat}\\
\delta p_\textrm{L} & =\frac{1}{2}R-\frac{m_{\textrm{C}^{\prime}}^{2}+\delta\vec{p}_\textrm{T}^{\,2}}{2R},\textrm{~with}\\
R 					& \equiv m_\textrm{C}+p_\textrm{L}^{\,\mu}+p_\textrm{L}^{\,\hadron}-E_\mu-E_\hadron,\textrm{~and finally}\\
\pN                 & = \sqrt{\delta p^2_\textrm{T} + \delta p^2_\textrm{L}}, \label{eq:pn}
\end{empheq}
where $p_{\kappa}$ is the momentum of particle $\kappa$. The subscripts    T and L stand for transverse and longitudinal components with respect to the \n direction (Fig.~\ref{fig:stkidiagram}). $\delta\vec{p}_\textrm{T}$ is the missing transverse momentum, or missing pT, between the final state $\mu+\hadron$ and the initial state $\nu_\mu$; its longitudinal counterpart, $\delta p_\textrm{L}$, depends on the muon and hadron energies, $E_\mu$ and $E_\hadron$, and the carbon-nucleus mass, $m_{\textrm{C}^{(\prime)}}$, before (after) the interaction:
\begin{align}
m_{\textrm{C}^{\prime}} = m_\textrm{C} - m_\textrm{n} + b, 
\end{align}
where $m_\textrm{n}$ is the neutron mass. The excitation energy $b$ is  $+28.7$~\mev~\cite{Furmanski:2016wqo,Lu:2019nmf}, which  affects  $m_{\textrm{C}^{\prime}}$ only at a per mil level and therefore has a negligible systematic impact. 

The transverse boosting angle, \dalphat [Eq.~(\ref{eq:dat})], describes the angular deviation of $\delta\vec{p}_\textrm{T}$ from the transverse momentum transfer, $-\vec{p}_\textrm{T}^{\,\mu}$. The configuration $\delta\alpha_\textrm{T}=0$ is only possible due to Fermi motion. Because Fermi motion is isotropic, events subject to Fermi motion and no other nuclear effects are evenly distributed in \dalphatn. Processes with intranuclear momentum transfer (IMT), such as final-state interactions (FSIs), pion absorption by the remnant, and production of extra particles included in X, will increase \dalphatn; the resulting non-flatness in the \dalphat distribution indicates how much impedance the outgoing hadrons experience inside the nucleus~\cite{Lu:2015tcr, Lu:2018stk, Abe:2018pwo}.

Following Eq.~(\ref{eq:pn}), it can be shown using a general event generator that \pn is approximated by \dpt with an $\mathcal{O}(10\%)$ correction. On the one hand, when X is carbon-11 in Eq.~(\ref{eq:topo}) and the final-state proton and neutral-pion do not experience FSIs, \pn is the initial momentum of the struck neutron in the resonant production: 
\begin{align}
    \nu_\mu + \neutron \to \mu^- +\proton + \pi^0.\label{eq:nucleonlevel}
\end{align}
On the other hand, in the presence of IMT, \dpt increases due to the extra momentum lost by FSIs or the unaccounted-for momentum share with other final-state particles in X, leading to a higher value of \pn beyond the Fermi surface.  

The initial neutron Fermi motion and IMT in QE-like interactions, including also 2-particle-2-hole (2p2h) contributions, have been measured and discussed in Ref.~\cite{Lu:2018stk}. This paper reports the first measurement of TKI(-based) final-state correlations, \dalphat and \pnn, in pion production. The measured Fermi motion and IMT dynamics will be compared to the QE-like measurement. 

\begin{figure}[!htb]
\centering
\includegraphics[width=\sigwidth]{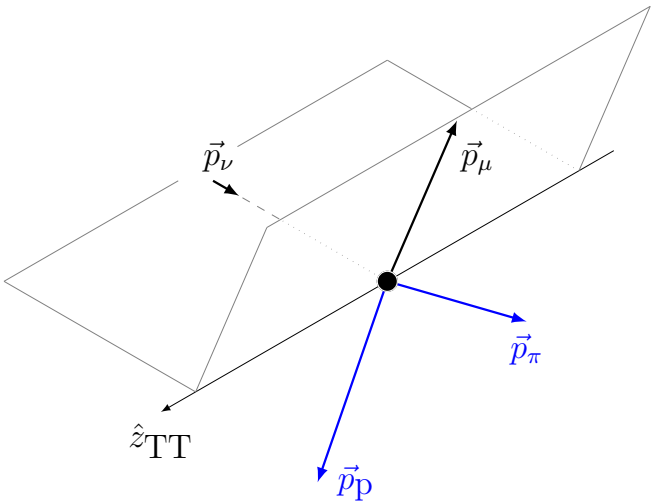}
\caption{Schematic diagram of the double-transverse kinematics~\cite{Lu:2015hea}. The neutrino and muon momenta define the double-transverse axis $\hat{z}_\textrm{TT}\equiv\vec{p}_\nu\times\vec{p}_\mu/|\vec{p}_\nu\times\vec{p}_\mu|$. The scattering plane is spanned by the lepton momentum vectors and therefore perpendicular to $\hat{z}_\textrm{TT}$. }\label{fig:dtkidiagram}
\end{figure}

In addition to studying the hadronic final state as a whole, the  proton and pion final states can be compared against each other to gain new insights into the hadronic momentum configuration in pion production as follows: the double-transverse momentum imbalance, \dpttn, describes the momentum imbalance of the final-state hadrons along the axis that is perpendicular to the lepton scattering plane~\cite{Lu:2015hea}:
\begin{align}
\delta p_\textrm{TT} & = \delta\vec{p}_\textrm{T} \cdot\hat{z}_\textrm{TT} \label{eq:dptdptt} \\
& = \vec{p}_\hadron\cdot\hat{z}_\textrm{TT},
 \end{align}
  where $\hat{z}_\textrm{TT}$ is a unit vector along $\vec{p}_\nu\times\vec{p}_\mu$ (Fig.~\ref{fig:dtkidiagram}). In the absence of nuclear effects, \dpt vanishes and so does  \dpttn, whereas for  nuclei with $\textrm{A}>1$, non-zero contributions arise from the nuclear medium. Because the proton and pion spatial distributions in resonant production might be asymmetric with respect to the lepton scattering plane (cf. Ref.~\cite{Cai:2019jzk} and references therein), any particle-type-dependent nuclear effects could cause an asymmetric distribution of \dpttn. This paper presents the first measurement of this new TKI.

\section{Measurement}

In the present work, the signal is defined as CC \numu events on carbon whose final state is \topo with any number of neutrons and no other particles exiting the nucleus [Eq.~(\ref{eq:topo})].
Kinematic constraints, resulting from detector acceptance and response,   are placed on the final state satisfying
\begin{empheq}[left=\empheqlbrace]{align}
	1.5 \leq p_\mu~(\mathrm{GeV}/c) \leq 20.0,~\theta_\mu < 25^\circ,\label{eq:mupscons}\\
	p_\textrm{p} \geq 0.45~\mathrm{GeV}/c,\label{eq:prpscons}
\end{empheq}
where  $\theta_\mu$ is the muon polar angle with respect to the \n direction. The  momenta of the leading proton and \piz are used in the calculation of the observables. 

The analysis uses data obtained with the \Min  detector
exposed to the NuMI low energy neutrino beam ($\langle E_\nu \rangle = 3~\mathrm{GeV}$) with
$3.33\times10^{20}$ protons on target (POT). The total neutrino flux ($2.88\times10^{-8}/\mathrm{cm}^2/\mathrm{POT}$) is estimated
according to Ref.~\cite{Aliaga:2016oaz}. Neutrino interactions inside a fiducial volume within \Minn's active tracker with a mass of 5.3 tons are selected. 
Precise tracking is achieved inside this fiducial volume by an alternating arrangement of hexagonal plastic scintillator planes at $0^\circ$ and $\pm60^\circ$ to the vertical. Each plane consists of 127 triangular polystyrene (CH) strips up to 245 cm long, 1.7 cm in height and a width of 3.3 cm. The strips form a plane by alternating strips such that the cross sectional view is a regular trapezium. 
Located two meters downstream of \Min is the \minos near detector---a magnetised muon spectrometer used to measure both the charge and momentum of muons.
A detailed description of both detectors can be found in Refs.~\cite{Aliaga:2013uqz,Michael:2008bc}.  

\N events are simulated in the detector using \genien~\cite{Andreopoulos:2009rq} \mnvgver where the initial state is modeled as a relativistic global Fermi gas (RFG)~\cite{Bodek:1980ar}.
\genie describes CC QE processes following Ref.~\cite{LlewellynSmith:1971uhs}  with a dipole axial  mass ($M_\textrm{A}^\textrm{QE}$) of 0.99~\gevn~\cite{Bradford:2006yz}.
The production of $\Delta$ and higher resonances used the Rein-Sehgal single pion model~\cite{Rein:1980wg}. Non-resonant pion production and multi-pion resonance contributions are introduced with a \genien-specific model~\cite{Bodek:2004pc}.
This background component is simulated up to a hadronic invariant mass range of $W < 1.7$~\gevn~\cite{Gallagher:2006ab,Wilkinson:2014yfa,Rodrigues:2016xjj}.
All resonant baryons decay isotropically in their rest frame with the exception of the $\Delta^{++}$. 
Following~\cite{Eberly:2014mra}, the $\Delta^{++}$ angular isotropy is suppressed by 50\% of that predicted by Rein-Sehgal.
Deep inelastic scattering (DIS) is incorporated into \genie via the 2003 Bodek-Yang model~\cite{Bodek:2002ps} and hadronization is described by \pythia~\cite{Sjostrand:2006za} and models based on Koba-Nielsen-Olesen scaling~\cite{Yang:2009zx, Koba:1972ng}.

\genien's default simulation is augmented to incorporate recent developments in both theory and experimental results as follows: 
2p2h contributions based on the Valencia model~\cite{Nieves:2011yp, Sobczyk:2012ms, Gran:2013kda, Schwehr:2016pvn} are included; 
the relative strength has been scaled upwards in accord with the \Min low-recoil measurement~\cite{Rodrigues:2015hik}.
Long-range correlations are included in QE interactions via the random phase approximation~\cite{Nieves:2004wx}.
Finally a reduction of 53\% in \genien's non-resonant single-pion prediction is applied in accord with recent analyses of deuterium
bubble chamber data~\cite{Wilkinson:2014yfa,Rodrigues:2016xjj}.
\genie applies an effective model of \fsi based on Ref.~\cite{Dytman:2011zz}.


The propagation of final state particles within the detector is simulated  using  \geant \gfvern~\cite{Agostinelli:2002hh}.
Hadron test beam data provided by a scaled down version of \Min  are used to constrain the \geant simulation of protons and charged pions~\cite{Aliaga:2013uqz}.
For both data and simulation the energy scale is calibrated using through-going muons.
These procedures ensure that the energy deposited per plane agrees between data and simulation. 

Signal-like events are selected by first requiring a single track originating from within \Minn's tracker to match a negatively charged track identified by \minosn.
This track must fulfill the kinematic constraints outlined in Eq.~(\ref{eq:mupscons}).
The muon's starting position, or primary vertex, is assessed for the existence of any additional tracks. 
In instances where extra tracks exist the primary vertex is redetermined to account for the extra information provided by these tracks.
All non-muon tracks are required to be proton-like by comparing their measured $\textrm{d}E/\textrm{d}x$ profiles to the simulated ones for protons and charged pions. 
Only proton-like tracks  are retained and their ranges are used to determine their momenta.
The leading proton must pass the phase-space requirement from Eq.~(\ref{eq:prpscons}).

Neutral pions are identified from their dominant decay signature, $\pi^0\to\gamma\gamma$, by requiring exactly two electromagnetic showers.
Their direction must be consistent with originating from the primary vertex. 
The calorimetric energy and direction of both photons are combined to reconstruct the \pizn's momentum.
Any remaining charged pion background is reduced by requiring that no Michel electron-like signature (indicating the presence of final-state $\pi^+$s) exist in the candidate events. The signal purity is improved by reconstructing the invariant mass, $m_{\gamma\gamma}$, of the two photons using
\begin{equation}
	m_{\gamma\gamma} = \sqrt{2E_1E_2\left(1-\cos\theta_{12}\right)},
	\label{eq:mgg}
\end{equation}
where $E_1$ and $E_2$ are the photon energies, and $\theta_{12}$ is the opening angle between the two photons.
Signal events are required to be within $60\leq m_{\gamma\gamma}~(\textrm{MeV}/c)\leq 200$. Full details of the selection can be found in previous \Min measurement~\cite{Altinok:2017xua}  which, however, placed an upper bound on the (experimental) hadronic invariant mass $W$ at 1.8~GeV/$c^2$ and required there be one and only one $\pi^0$ regardless of the protons in the final state.

\begin{figure}[!htb]
\centering
\includegraphics[width=\columnwidth]{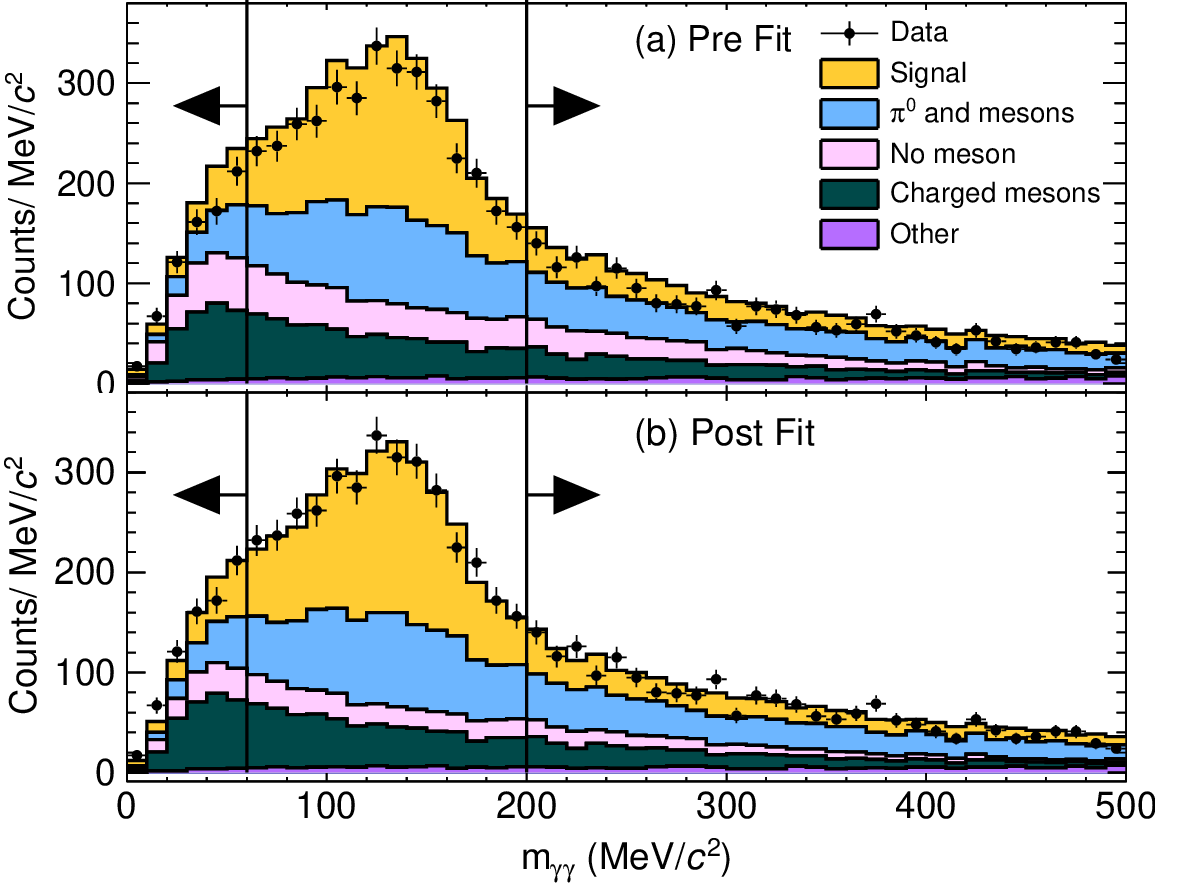}
\caption{Reconstructed distributions of the two-photon invariant mass in the selected sample, compared to simulations (a) before and (b) after the background fit. For completeness, the excluded regions are also shown, indicated by the arrows. The \textit{Other} category contains \topo events which are out of acceptance.}\label{fig:mggsig}
\end{figure}
\begin{figure}[!htb]
\centering
\includegraphics[width=\columnwidth]{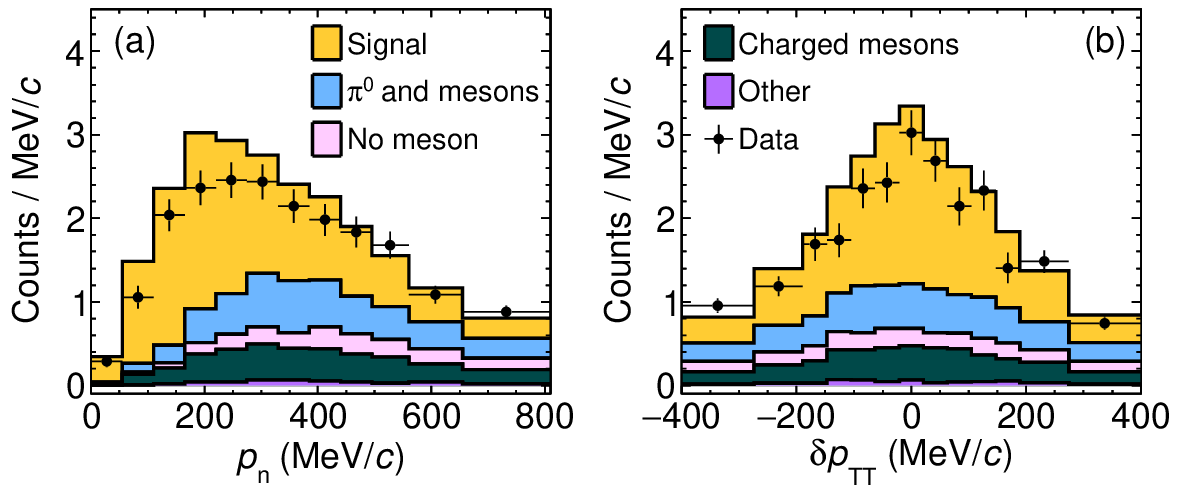}
\caption{Reconstructed distributions of (a) \pn and (b) \dpttn, compared to simulations after the background fit. }\label{fig:pNpTT_Reco}
\end{figure}

 The resulting sample has 51.4\% purity and 5.7\% efficiency. 
For a neutrino-proton interaction, the \topo final-state requirement only allows multiproton production which is highly suppressed. Therefore, almost all events from the hydrogen component of the CH target contribute to the background. 
Overall, the dominant background categories are (A) \piz events with other mesons (\piz \textit{and mesons}),  (B) events without \pizn s (\textit{Charged mesons}),  and (C) zero-meson events (\textit{No meson}), in decreasing order of importance, as is shown in Fig.~\ref{fig:mggsig}. 
A data-driven approach is used to constrain these background components. 
Three sidebands are obtained by loosing one of the signal selection criteria: one sideband utilized events below and above the $m_{\gamma\gamma}$  range allowed to the selected sample; the second one used
events that fail the quality requirements for proton tracks; the third one used events that were accompanied by
a Michel electron tag.  
The size of the backgrounds (A)-(C) is tuned to describe the data in these sidebands.  The resulting  scaling factors are 0.92, 1.12, and 0.67, respectively. Compared to Ref.~\cite{Altinok:2017xua}, all the backgrounds scale in the same way:  the charged-meson component (B) increases whereas the other two decrease. Because the signal definition in Ref.~\cite{Altinok:2017xua} is different, the scaling factors for backgrounds (A) and (C) are significantly updated as expected. Details of the background fit can be found in Ref.~\cite{Coplowe:2018fic}. The postfit distributions are shown in Fig.~\ref{fig:pNpTT_Reco} for the reconstructed \pn and \dpttn. 

The reconstructed proton momentum resolution is improved by selecting elastically scattered and contained protons via an additional  criterion that requires a large $\textrm{d}E/\textrm{d}x$ near the track end-point~\cite{Lu:2018stk}.
As a result this removes ill-determined momentum-by-range events whose selected protons either undergo inelastic scattering or are not contained in the tracker.
This leads to a $p_\textrm{p}$ resolution of $\sim2\%$ at 1~\gevn/$\textit{c}$ albeit at the cost of a 50\% reduction in statistics.

The \piz momentum reconstruction is improved via kinematic fitting~\cite{Baer:1980qf, Stroher:1988rj}. Given the relationship between the \piz mass and the photon kinematics in Eq.~(\ref{eq:mgg}), the photon energies are recalculated by minimizing
\begin{equation}
	\chi^2 = \left[\frac{m^2_\pi - 2E_1E_2\left(1-\cos\theta_{12}\right)}{\sigma_{m^2_\pi}}\right]^2 + \sum_{i=1}^2\left[\frac{E_i - E_{i}'}{\sigma(E_i)}\right]^2,
\end{equation}
where $E_{i}'$ are the measured photon energies, and the recalculated energies, $E_{i}$, are treated as free parameters in the fit. 
The second term acts as a penalty for each photon and ensures that the fitted energies are within expectation of their calorimetrically measured values. 
Note that $\sigma_{m^2_\pi}$, representing the resolution of the reconstructed $\pi^0$ mass, is used as an optimisation parameter whose value is chosen such that 99\% of the fits successfully converge.
The photon energy resolution, $\sigma(E_i)$, is determined from simulation. 
A full description can be found in Ref.~\cite{Coplowe:2018fic}. This leads to a  $\pi^0$ momentum resolution of about 20\%.


Flux-integrated cross sections are produced by first subtracting the constrained backgrounds from the selected samples.
D'Agostini unfolding~\cite{DAgostini:1994fjx} is then performed with 4 iterations. The unfolding procedure is validated by reproducing pseudodata that is generated by extreme variations of the cross section models. The efficiency correction is then applied, followed by event normalization by the product of the flux and number of target nucleons (\ntargets).
Systematic uncertainties are evaluated for all observables following Ref.~\cite{Altinok:2017xua}. In particular, parameters in the physics and detector models are varied within uncertainties and the resulting cross section variations are the assigned systematic uncertainties. For example, \pnn, whose statistical uncertainty spans 10--34\%, has systematic uncertainties arising from detector (2--8\%), flux (3--8\%), and \genie cross section models (5-28\%); as one of the \genie model parameters, the aforementioned $M_\textrm{A}^\textrm{QE}$ leads to an uncertainty of 0.1-1\%. The total uncertainty for \pn at few~MeV/\textit{c} is approximately 22\%, increasing to 46\% at 0.8~GeV/\textit{c}. (See Supplemental Materials 1~\cite{supp1} and 2~\cite{supp2} for details of cross-section uncertainties.)

\section{Results}

\begin{figure}[!htb]
\centering
\includegraphics[width=\sigwidth]{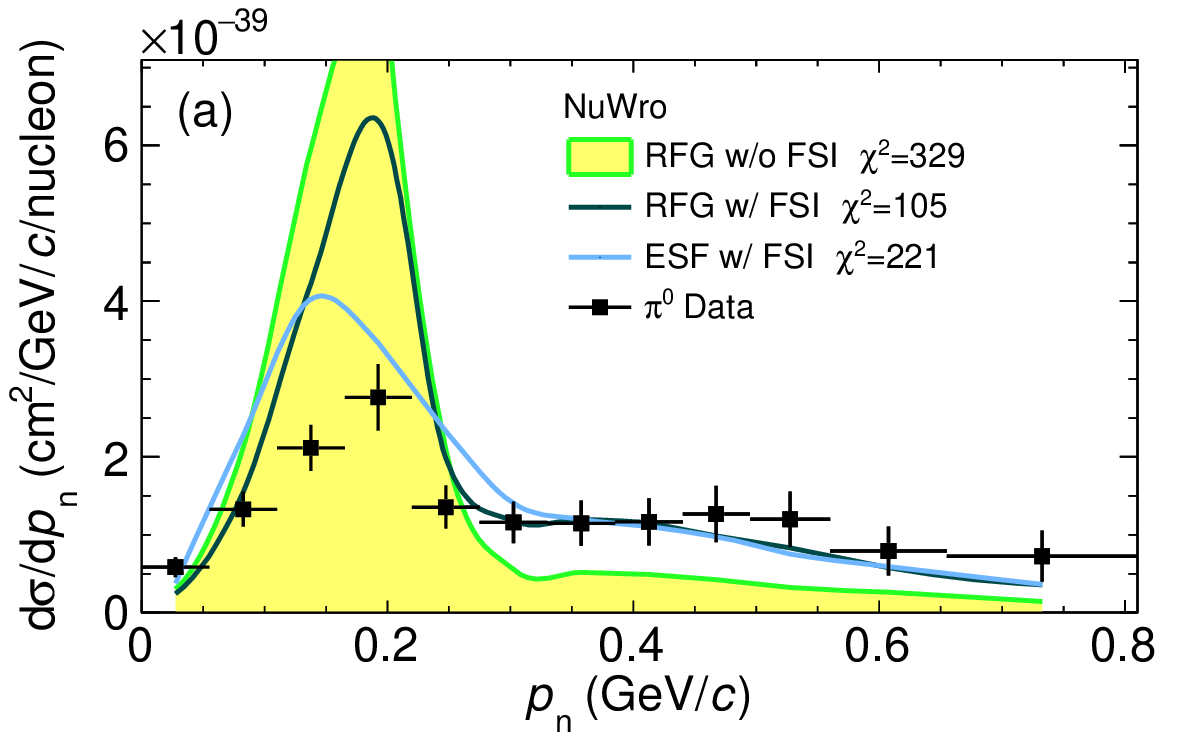}
\includegraphics[width=\sigwidth]{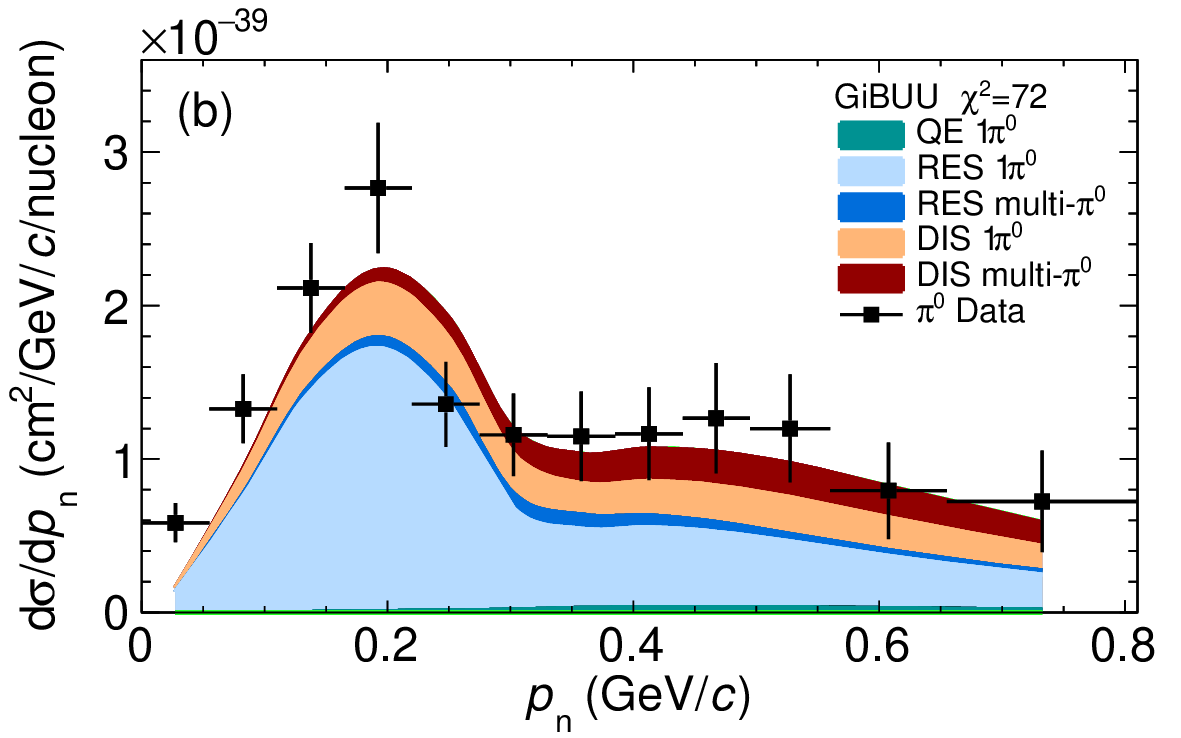}
\caption{Cross section in \pn compared to (a) \nuwro \nver and (b) \gibuu \bver predictions. Error bars on the data  include both statistical and systematic uncertainties. The \nuwro prediction for RFG without \fsi has a maximum of $8.9\times10^{-39}~\textrm{cm}^2/\textrm{GeV/}c/\textrm{nucleon}$. The GiBUU predictions are decomposed into single-and multi-$\pi^0$ contributions.}\label{fig:pN_NuWroIS}
\end{figure}

The measured cross section in \pn is compared to generator predictions in Fig.~\ref{fig:pN_NuWroIS}.
The Fermi motion peak  (below 0.25~GeV/$c$) is qualitatively captured by the \nuwro (\nvern)~\cite{Golan:2012wx} RFG model.  In this Fermi gas model, all nucleons lie below  the Fermi surface and the predicted cross section in \pn has a cut-off at 0.22~GeV/\textit{c}. 
For the previous QE-like measurement, the Spectral Function (SF) approach~\cite{Benhar:1994hw} best describes the data~\cite{Lu:2018stk}. However, at present,  while SF calculations for  pion production  exist~\cite{Nakamura:2007pj, Nakamura:2008zzd}, they are not yet implemented in generators. In \nuwron, the Effective Spectral Function (ESF)~\cite{Ankowski:2005wi}  incorporates the most important features of SF in generator implementation: in ESF, the probability distribution of the target nucleon momentum is identical to SF; for a selected value of the nucleon momentum, an average removal energy calculated from SF is used. 

The non-exclusive part of the signal, such as multi-$\pi^0$ production, gives rise to large values of \pn beyond the Fermi surface, and hence the long tail in the \nuwro RFG prediction without FSIs. When FSIs are switched on, kinematic distortion migrates  events away from the Fermi motion peak; pion  absorption and charge exchange following multi-$\pi$ contributions and wrong-sign ($\Delta^{++}$) production also add to the tail region. These IMT processes lead to a \pn tail that is similar in the RFG and ESF predictions. 
In this large-missing-pT region, \nuwro with either initial-state models describes data within about 1-$\sigma$. However, in the peak region, the data exhibit a distinctly muted distribution devoid of the sharp falloff.  Furthermore, the ESF peak locates at around 0.15~GeV/\textit{c}, 25\%  off compared to data. 

Comparison is also made to \gibuu (\bvern)~\cite{Leitner:2006ww, Leitner:2008ue, Buss:2011mx} predictions in Fig.~\ref{fig:pN_NuWroIS} (b). While it also describes the large-missing-pT region, \gibuu underpredicts the Fermi motion peak. Nevertheless, it has the correct peak location and overall better describes  the data. In \gibuun, the initial state is modeled as local Fermi gas in a  nuclear potential~\cite{Mosel:2019vhx}.
Model features that decrease or enhance the exclusive proton-$\pi^0$ production will have as large an effect on the agreement as the initial state.    
From the decomposition of the interaction modes, 
it can be seen that besides the dominant resonant production, the DIS has a sizeable contribution. Furthermore, the DIS contribution to the Fermi motion peak is dominated by single-$\pi^0$ production, and QE events wherein proton FSI initiates $\pi^0$ production give a small contribution.
 
\begin{figure}[!htb]
\centering
\includegraphics[width=\sigwidth]{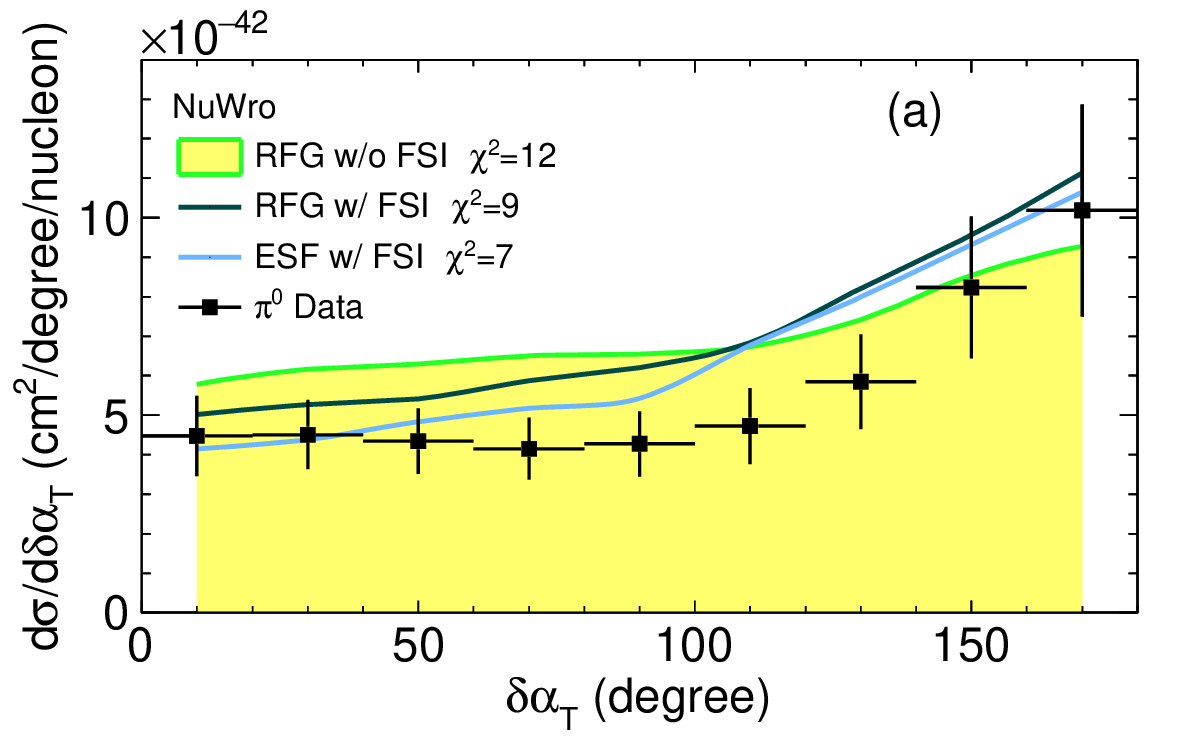}
\includegraphics[width=\sigwidth]{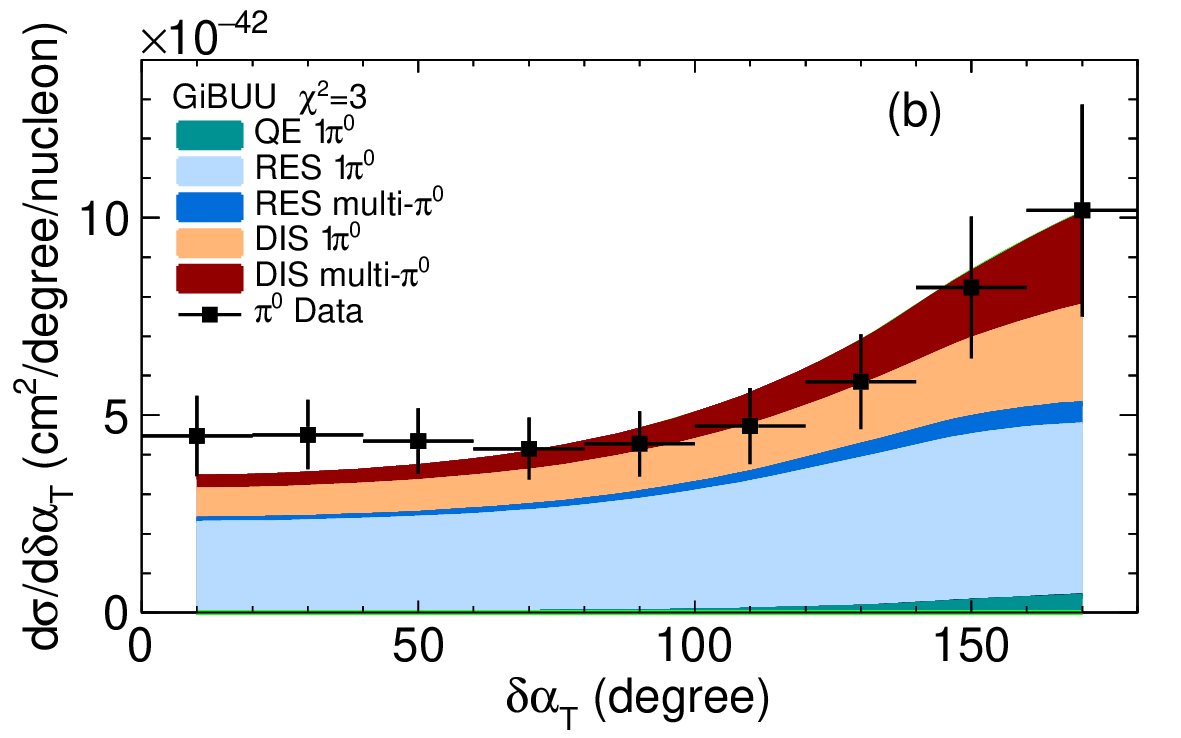}
\caption{Cross section in \dalphat compared to (a) \nuwro \nver and (b) \gibuu \bver predictions.}\label{fig:dat_NuWroIS}
\end{figure}

The measured cross section in \dalphat is shown in Fig.~\ref{fig:dat_NuWroIS} with model predictions. Because Fermi motion is isotropic, events in the \pn peak are evenly distributed in \dalphat and therefore provide a flat baseline for the overall cross section. The slope of the cross section towards $\delta\alpha_\textrm{T}=180^\circ$ comes from IMT events in the \pn tail. For the prediction without FSIs, the non-exclusive part of the signal leads to the rise in cross section at large \dalphatn. With FSIs, the \nuwro predictions become steeper and are much more similar with each other than in the case of \pnn---this is expected as \nuwro FSI is decoupled from the initial state that gives a model-independent flat baseline. 
As \nuwro overpredicts the \pn peak size with RFG and ESF, the overall predictions for \dalphat are above the data. 
For \gibuun, the slope is similar to that for \nuwro but the overall agreement is better because of the lower Fermi motion baseline.  

\begin{figure}[!htb]
\centering
\includegraphics[width=\sigwidth]{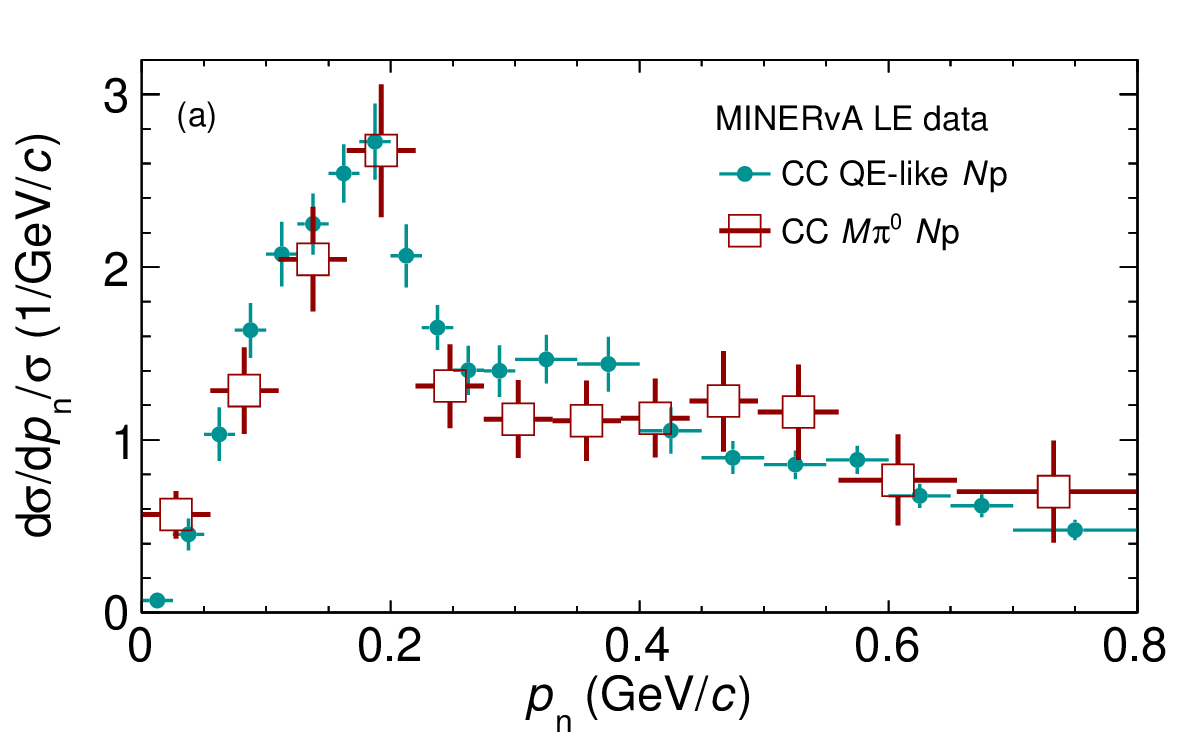}
\includegraphics[width=\sigwidth]{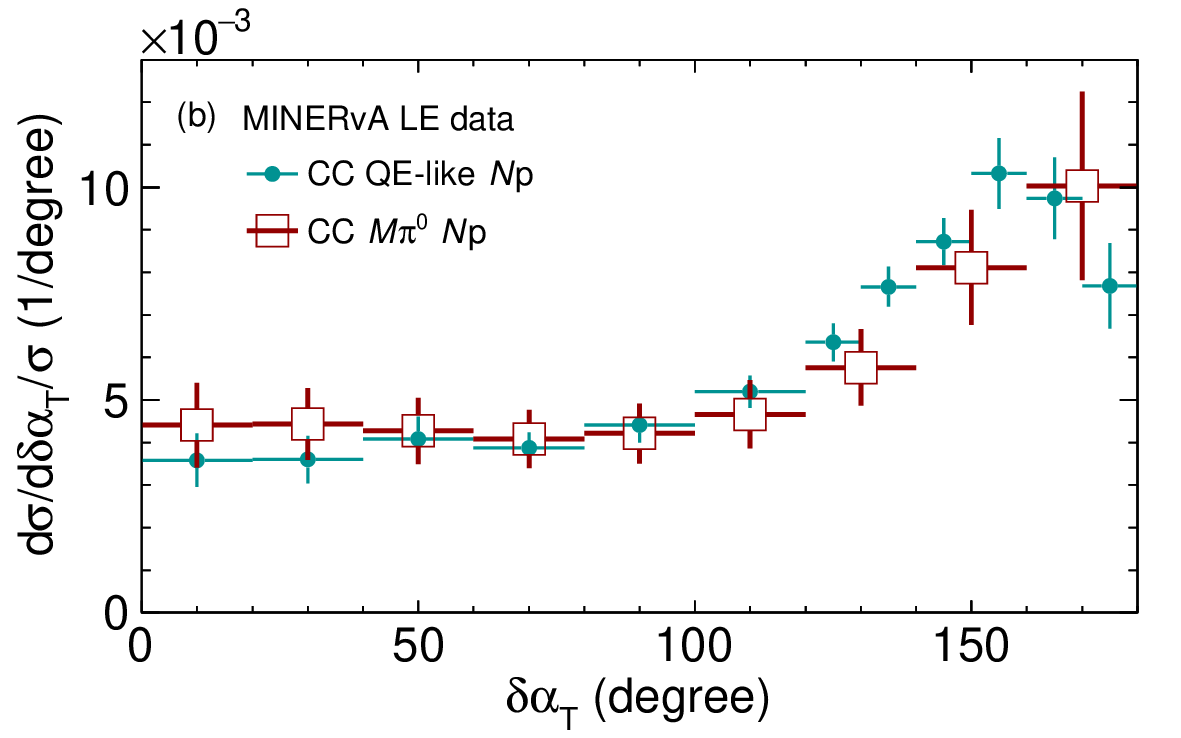}
\caption{Area-normalized cross sections in (a) \pn and (b) \dalphatn. Results from this work ($M\pi^0 N\proton$) are compared to the previous \Min CC QE-like measurement~\cite{Lu:2018stk}. Both measurements are based on the same data set using the NuMI low energy (LE) neutrino beam.}\label{fig:pi00pi}
\end{figure}

In Fig.~\ref{fig:pi00pi}, the cross sections in \pn and \dalphat are area-normalized and compared to those from the previous \Min CC QE-like measurement~\cite{Lu:2018stk}:
\begin{equation}
	\nu_\mu + \textrm{C} \to \mu^- +\proton +\textrm{X}',
	\label{eq:QEtopo}
\end{equation}
where X$^\prime$ is a final-state hadronic system consisting of the nuclear remnant with possible additional protons but without mesons. Both measurements are based on the same data set using the NuMI low energy neutrino beam. Similar to the discussions for Eqs.~(\ref{eq:topo}) and~(\ref{eq:nucleonlevel}), in Eq.~(\ref{eq:QEtopo}) when X$^\prime$ is carbon-11 and the final-state proton do not experience FSIs, \pn is the initial momentum of the struck neutron probed in the QE scattering:
\begin{align}
\nu_\mu + \neutron &\to \mu^- +\proton. \label{eq:QEnucleonlevel}
\end{align}
Comparing the shape of the \pn Fermi motion peak between the two measurements in Fig.~\ref{fig:pi00pi} (a), consistency is expected as the same initial state is probed. It is the consistency in the tail size (relative to the peak) that is nontrivial. In the large-missing-pT region, the underlying IMT processes are different: FSIs in QE, pion absorption in resonant production, and 2p2h on the one hand for Eq.~(\ref{eq:QEtopo}) (see Ref.~\cite{Lu:2018stk} and Supplemental Material 1~\cite{supp1}), FSIs in resonant production, and DIS with or without pion absorption on the other hand for Eq.~(\ref{eq:topo}). Most notably, pion absorption increases the tail size in QE-like events, but decreases it in single-pion production. 

The dynamics in the large-missing-pT region is further analyzed in the comparison of \dalphat in Fig.~\ref{fig:pi00pi} (b). 
Since the slope towards large \dalphat indicates how much impedance the out-going hadrons experience inside the nucleus, the similar slopes in both measurements seem to suggest another possible fine tuning: similar impedance in both QE-like and pion production, which is surprising because of the very different underlying IMT processes. While the consistency in both the \pnn-tail size and the large-\dalphat slope seem to be accidental, a combined analysis of both samples could provide very precise constraints on the modeling of the IMT processes. For completeness, \gibuu predictions for the QE-like measurement can be found in Supplemental Material 1~\cite{supp1}.

\begin{figure}[!ht]
\centering
\includegraphics[width=\sigwidth]{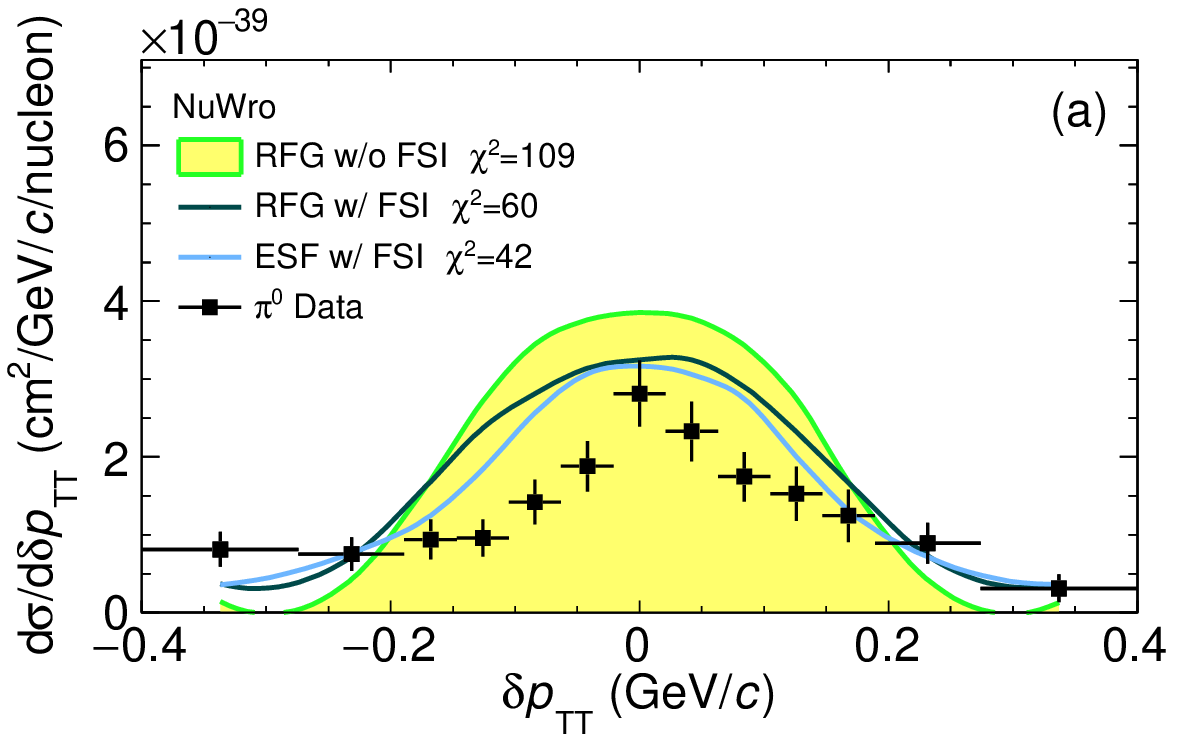}
\begin{overpic}[width=\sigwidth]{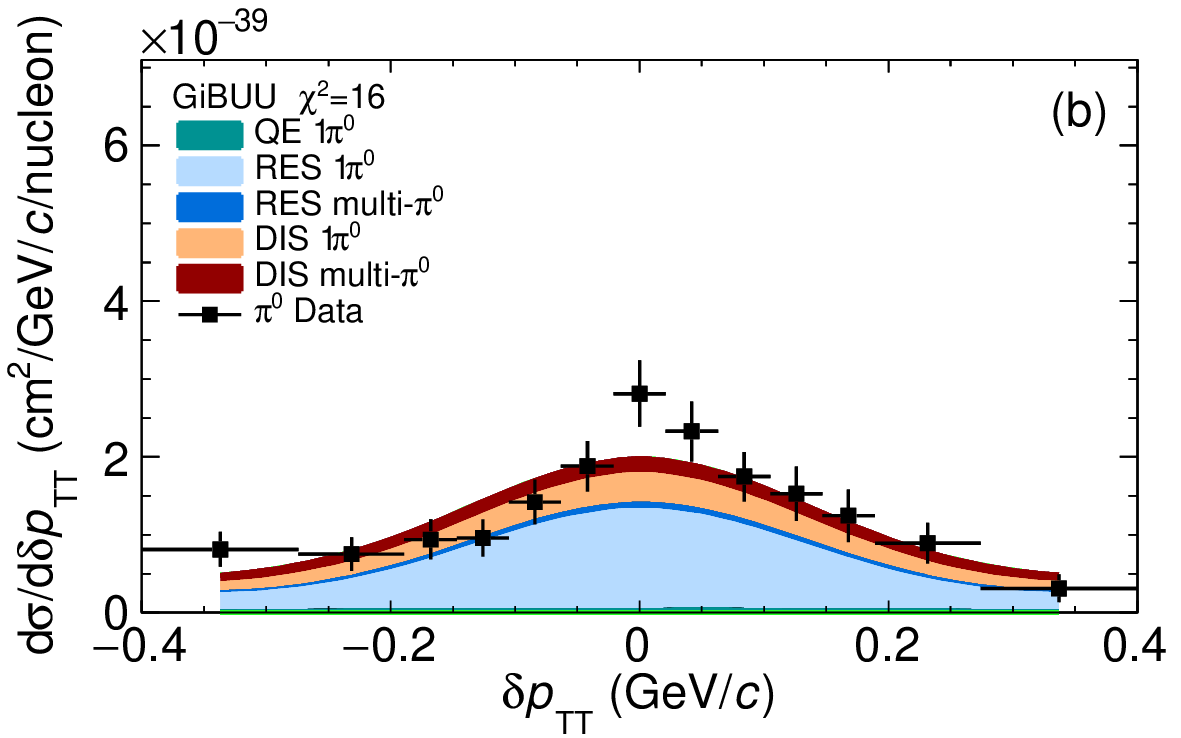}
\put(118,80){\includegraphics[width=0.38\columnwidth]{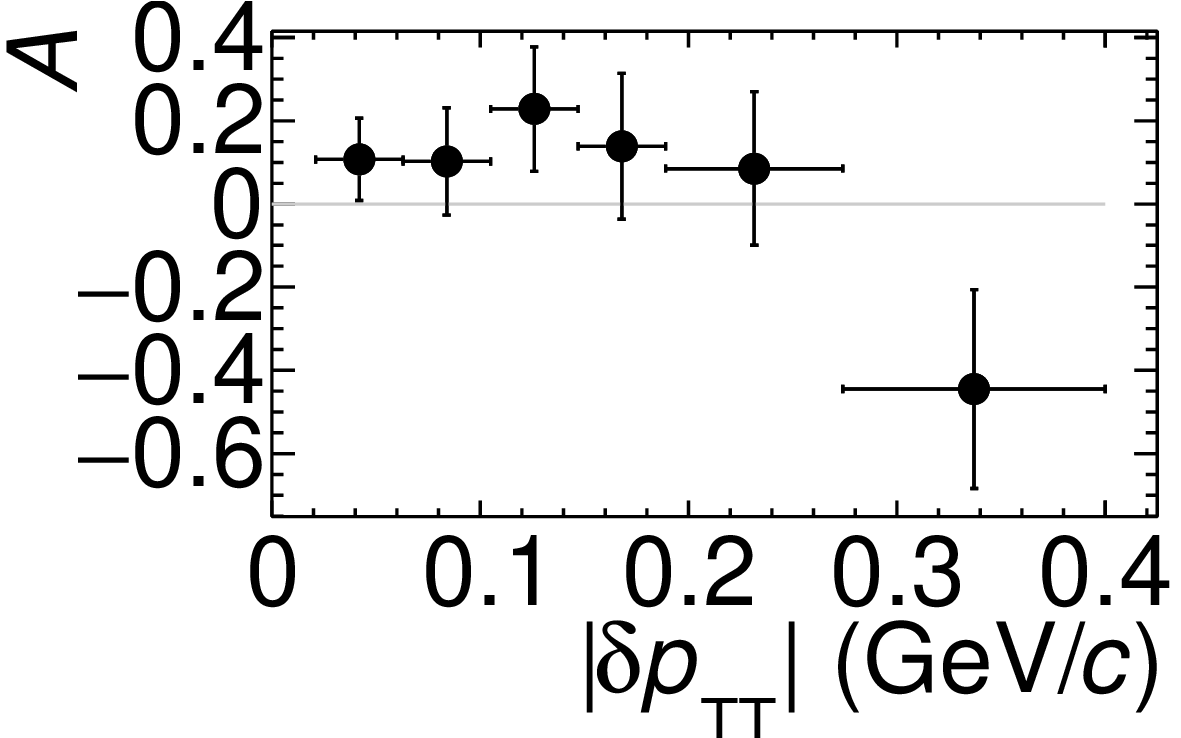}}
\end{overpic}
\caption{Cross section in \dptt compared to (a) \nuwro \nver and (b) \gibuu \bver predictions. The \dptt  asymmetry, $A$ from Eq.~(\ref{eq:asydef}), is shown as a function of $\left|\delta p_\textrm{TT}\right|$ in the inset in (b).  }\label{fig:pN_GENIEIS}
\end{figure}

The cross section in \dptt is shown in Fig.~\ref{fig:pN_GENIEIS}, compared to \nuwro and \gibuu predictions. According to Eq.~(\ref{eq:dptdptt}), part of the data-model discrepancy can be traced back to the mismodeling at the \pn peak as discussed above. The new aspect of nuclear effects probed by \dptt is the symmetry of the cross-section shape with respect to $\delta p_\textrm{TT}=0$. Without nuclear effects, the proton and $\pi^0$ momenta from Eq.~(\ref{eq:nucleonlevel}) would be balanced with respect to the lepton scattering plane, yielding $\delta p_\textrm{TT}=0$. Fermi motion and IMT contributions then introduce event-by-event imbalance to either side of the plane. Previous measurements including Ref.~\cite{Altinok:2017xua} indicate that the final-state $\pi^0$ could prefer one side of the scattering plane due to the interference between Delta and non-resonant amplitudes (see Ref.~\cite{Cai:2019jzk} for discussions). Without loss of generality, suppose the proton in Eq.~(\ref{eq:topo}) would always prefer the positive-$\hat{z}_\textrm{TT}$ side of the lepton scattering plane (Fig.~\ref{fig:dtkidiagram}). In such a case, we would have $\delta p_\textrm{TT}=|\vec{p}_\proton\cdot\hat{z}_\textrm{TT}|-|\vec{p}_{\pi^0}\cdot\hat{z}_\textrm{TT}|$. FSIs that change the proton and \piz momenta differently could then cause a positive-negative asymmetry in \dpttn. In this measurement, the cross-section asymmetries of each positive-negative \dpttn-bin pair are shown in the Fig.~\ref{fig:pN_GENIEIS} (b) inset. They are defined as
\begin{align}
A\left(\left|\delta p_\textrm{TT}\right|\right)\equiv\frac{\textrm{d}\sigma\left(\left|\delta p_\textrm{TT}\right|\right) - \textrm{d}\sigma\left(-\left|\delta p_\textrm{TT}\right|\right)}{\textrm{d}\sigma\left(\left|\delta p_\textrm{TT}\right|\right) + \textrm{d}\sigma\left(-\left|\delta p_\textrm{TT}\right|\right)}.\label{eq:asydef}
\end{align}
Compared to these observed mild asymmetries, negligible asymmetry is predicted by the generators. 

\section{Summary}

This paper presents the first measurements of a set of novel final-state correlations in neutrino CC $\pi^0$ production. These TKI-based correlations include the transverse boosting angle (\dalphatn) and the emulated nucleon momentum (\pnn), both of which were previously measured only in CC QE-like production~\cite{Lu:2018stk, Abe:2018pwo}, as well as the double-transverse momentum imbalance (\dpttn) that has no previous measurement. This work separates Fermi motion and IMT in pion production for the first time, and the observed cross-section shapes are consistent with the previous \Min measurement of QE-like interactions. 
In the present measurement,  RFG and ESF models   describe within 1-$\sigma$ the  large-missing-pT region (\pnn$>300$~MeV/\textit{c}) that collectively comes from IMT processes including \fsi distortions,  non-exclusive contributions, and wrong-sign ($\Delta^{++}$) production. 
The measured exclusive proton-$\pi^0$  production cross section on neutrons in nuclei in the Fermi motion peak region (\pnn$<220$~MeV/\textit{c}), however, is mismodeled by current generators using Fermi gases. Whereas Spectral Function can successfully describe the previous QE-like results, the Effective Spectral Function approach still overpredicts the Fermi motion peak in pion production. New investigation of the final-state hadronic momentum configuration is made with \dpttn, showing a mild asymmetry that could come from particle-type-dependent FSIs following interference effects in pion production. 

In the future, nuclear effects associated with initial-state protons could be examined using proton-$\pi^\pm$ production in CC $\smash{\overset{\scalebox{.3}{(}\raisebox{-1.7pt}{--}\scalebox{.3}{)}}{\nu}_\mu}$-nucleus scattering~\cite{Lu:2019nmf}.   Measurements of proton-$\pi^0$ systems as reported here could be performed in liquid argon TPC experiments such as ICARUS~\cite{Antonello:2015lea}, MicroBooNE~\cite{Adams:2018sgn},  and SBND~\cite{Antonello:2015lea}, providing clarifications of neutrino-argon  interactions that are needed by the DUNE neutrino oscillation program.  \Minn’s medium-energy exposures allow new investigations at higher neutrino  
energies and with larger event samples~\cite{Valencia:2019mkf, Carneiro:2019jds}.   Thus,  further illumination of the  physics that underwrites the TKI-based final-state correlations in CC pion production  may be anticipated.

\begin{acknowledgments}
We thank Luis Alvarez Ruso for helpful comments on the manuscript. 
This document was prepared by members of the \Min Collaboration using the resources of the Fermi National Accelerator Laboratory (Fermilab), a U.S. Department of Energy, Office of Science, HEP User Facility. Fermilab is managed by Fermi Research Alliance, LLC (FRA), acting under Contract No. DE-AC02-07CH11359.
These resources included support for the \Min construction project, and support
for construction also
was granted by the United States National Science Foundation under
Award No. PHY-0619727 and by the University of Rochester. Support for
participating scientists was provided by NSF and DOE (USA); by CAPES
and CNPq (Brazil); by CoNaCyT (Mexico); by Proyecto Basal FB 0821, CONICYT PIA ACT1413, Fondecyt 3170845 and 11130133 (Chile); 
by CONCYTEC (Consejo Nacional de Ciencia, TecnologÌa e InnovaciÛn TecnolÛgica), DGI-PUCP (DirecciÛn de GestiÛn de la InvestigaciÛn  - Pontificia Universidad CatÛlica del Peru), and VRI-UNI (Vice-Rectorate for Research of National University of Engineering) (Peru);
and by the Latin American Center for Physics (CLAF); NCN Opus Grant No. 2016/21/B/ST2/01092 (Poland); by Magdalen College Oxford and Science and Technology Facilities Council (UK).  
We thank the MINOS Collaboration for use of its near detector data. Finally, we thank the staff of
Fermilab for support of the beam line, the detector, and computing infrastructure.
\end{acknowledgments}

\bibliography{main}

\end{document}